\documentstyle[onecolumn,epsfig]{mn}
\begin{document}

\title[Self-similar structure of disks]{Self-similar structure of the magnetized radiation-dominated accretion disks}

\author[Shadmehri and  Khajenabi]
{M. Shadmehri\thanks{E-mail:
mshadmehri@science1.um.ac.ir} and F. Khajenabi\thanks{E-mail:fkhajenabi@science1.um.ac.ir}\\
Department of Physics, School of Science, Ferdowsi University,
Mashhad, Iran}

\maketitle

\date{Received 10 April 2005 / Accepted _________________ }
%
%
\markboth{Shadmehri & Khajenabi: Self-similar structure of
disks}{}
\begin{abstract}
We investigate the effects of a large-scale magnetic field with
open field lines on the steady-state structure of a
radiation-dominated accretion disk, using self-similarity
technique. The disk is supposed to be turbulent and possesses an
effective viscosity and an effective magnetic diffusivity. We
consider the extreme case in which the generated energy due to
viscous and magnetic dissipation is balanced by the advection
cooling. While the magnetic field outside of the disk is treated
in a phenomenological way, the internal field is determined
self-consistently. Magnetized and nonmagnetized solutions have
the same radial dependence, irrespective of the values of the
input parameters. Generally, our self-similar solutions are very
sensitive to the viscosity or diffusivity coefficients. For
example, the density and the rotation velocity increase when the
viscosity coefficient decreases. The gas rotates with
sub-Keplerian angular velocity with a factor less than unity
which depends on the magnetic field configuration. Magnetic field
significantly reduce disk thickness, however,  tends to increase
the radial velocity comparing to the nonmagnetic self-similar
solutions.
\end{abstract}

\begin{keywords}
accretion, accretion disks - black hole physics - MHD
\end{keywords}
%

\section{Introduction}

Advection-dominated accretion flows (ADAFs) have been studied by
many authors during recent years (e.g., Abramowicz et al. 1988;
Narayan \& Yi 1994; Chen 1995; Narayan, Kato \& Honma 1997). A
key feature of radiatively inefficient accretion flows is that
radiative energy losses are small so that most of the energy is
advected with the gas. However, advection-dominated accretion
flows can occur in two regimes, depending on the accretion rate
and the optical depth. When the accretion rate is high, the
optical depth becomes very high and the radiation is trapped in
the gas. These kinds of solutions which are known under the name
'slim accretion disk' have been studied in detail by Abramowicz
et al. (1988). On other hand, we may have optically thin
accretion flows with very low mass accretion rate (e.g., Rees et
al. 1982; Narayan \& Yi 1994; Abramowitz et al. 1995; Chen 1995).

Because of the complexity of the equations, similarity technique
can help us to explore the relevant physics of radiatively
inefficient accretion flows. As long as we are not interested in
the boundaries of the problem, such solutions that describe the
behavior of the flow in an intermediate region far from the radial
boundaries. Originally, Narayan \& Yi (1994) studied optically
thin advection-dominated accretion disk using their self-similar
solutions. They speculated, on the basic of some numerical
calculations, that the self-similar solution is the natural state
for an advection-dominated flow. Subsequent analysis (e.g.,
Narayan, Kato \& Honma 1997) showed that the global solutions
achieve approximate self-similar behavior within a short distance
from the outer boundary and the approach to self-similarity is
quite impressive. Considering these achievements, Wang \& Zhou
(1999) constructed self-similar solutions for optically {\it
thick} advection dominated accretion flow, in which photon
trapping and advection dominate over surface diffusion cooling and
used these to explore its different properties from optically
thin self-similar solutions. However, both solutions have the same
power index of radius.

A remarkable problem arises when the accretion disk is threaded by
magnetic field. There are good reasons for believing that magnetic
fields are important to the physics of accretion disk. Schwartzman
(1971) was the first to point out the importance of the magnetic
field in an accretion process. He proposed a hypothesis of
equipartition between the magnetic and kinetic energy densities
and this picture  is usually accepted in the modern picture of
viscous ADAF models. Bisnovatyi-Kogan \& Lovelace (2000) suggested
that recent papers discussing ADAF as a possible solution for
astrophysical accretion should be treated with caution,
particularly because of ignorance of the magnetic field. While
they obtained a solution for a time-averaged magnetic field in a
quasispherical accretion flow, an analysis of energy dissipation
and equipartition between magnetic and flow energies has been
presented (Bisnovatyi-Kogan \& Lovelace 2000). Numerical
simulations of {\it magnetized}, radiatively  inefficient flows
have been done recently by several authors (e.g., Machida,
Matsumoto \& Mineshige 2001;  Igumenshchev, Narayan \& Abramowicz
2003). However, in most of these the resistive terms in the MHD
equations  have been neglected, or the resistivity has been
considered only in the induction equation without accounting the
corresponding dissipation in the energy equation.

Some authors tried to study magnetized accretion flow
analytically. For example, Kaburaki (2000) presented a set of
analytical solutions for a fully advective accretion flow in a
global magnetic field and the conductivity is assumed to be
constant for simplicity. Shadmehri (2004) extended this analysis
by considering non-constant resistivity. He obtained a set of
self-similar solutions in spherical coordinates that describes
quasi-spherical magnetized accretion flow. Lai (1998) and Lee
(1999) calculated transonic disk accretion flows around a weekly
magnetized neutron star, where it was assumed the disk is fully
advective. While Lee (1999) considered both the thermal and the
radiation pressures, Lai (1998) assumed the radiation pressure
dominates over the gas pressure.

In this study, we present self-similar solutions of an idealized
height-integrated set of equations that describe magnetized
radiation-dominated accretion flow. In fact, this analysis extends
self-similar solutions of  Wang \& Zhou (1999) for optically thick
advection-dominated accretion flow to the magnetized case. For a
steady state disk, there can be a final, steady configuration of
magnetic field, in which the inward dragging of field lines by the
disk is balanced everywhere by the outward movement of filed lines
due to magnetic diffusivity. We show that the radial structure of
a magnetized radiation-dominated accretion flow does not differ
from the non-magnetic solutions. But we can see significantly
different behaviors, because of the effects of the magnetic
fields. The equations of the model are presented in the second
section. We obtain and solve the set of self-similar equations
analytically in the third section. For a set of illustrative
parameters the solutions will be discussed  in this section.

\section{General Formulation}
We employ a cylindrical coordinate system $(R,\varphi,z)$ centered
on a central object (e.g., a black hole) with mass $M_{\ast}$
which accretes matter at a steady state  $\dot{M}$ from a
geometrically thin axisymmetric accretion disk in steady state
threaded by an ordered magnetic field. Our model generalize the
usual slim disks around black holes (e.g., Muchotrzeb \&
Paczy\'{n}ski 1982; Matsumoto et al. 1984; Abramowicz et al. 1988)
by including the effect of magnetic fields. General relativistic
effects are neglected and outside of the disk, dissipative effects
are assumed to be negligible. For the magnetic field geometry, we
are following a general approach presented by Lovelace, Romanova
\& Newman 1994 (hereafter LRN), in which even filed symmetry is
assumed so that $B_{\rm R}(R,z)=-B_{\rm R}(R,-z)$,
$B_{\varphi}(R,z)=-B_{\varphi}(R,-z)$ and $B_{\rm z}(R,z)=+B_{\rm
z}(R,-z)$. However, we note that the solution for the magnetic
field outside of disk should match to the field solution inside
the disk at its surface. But our model does not present a
self-consistent model for the magnetic field outside of the disk.
In analogy to LRN, we parameterize the magnetic field outside of
the disk. Also, it is assumed that the accreting matter is
confined to a thin disk, and we do not formally introduce a
magnetosphere into our model.

The basic equations are integrated over the vertical thickness of
the disk. The mass continuity equation takes the form
\begin{equation}
-2\pi R \Sigma v_{\rm R}=\dot{M},\label{eq:masscon}
\end{equation}
where $v_{\rm R}$ is the radial velocity and $\Sigma=\int dz \rho
\simeq 2 h \rho $ is surface density of the disk. The
half-thickness is denoted by $h$, where we consider the magnetic
field effect on the disk thickness. We will see that the disk can
be compressed or flattened depending on the field configuration.
Also, we note that since the radial velocity is negative for
accretion (i.e., $v_{\rm R}<0$), the accretion rate $\dot{M}$ as
an input parameter of our model is positive.

The radial momentum equation reads
\begin{equation}
\Sigma v_{\rm R}\frac{d v_{\rm R}}{d
R}-\Sigma\frac{v_{\varphi}^2}{R}=-\frac{d P}{d
R}-\Sigma\frac{GM_{\ast}}{R^2}+\int F_{\rm R}^{\rm mag}
dz,\label{eq:rcom}
\end{equation}
where $P=\int dz p$ is the integrated disk pressure. We consider
an extreme case in which the radiation pressure dominates over
the gas pressure and the generated energy is balanced by the
advection cooling. The first assumption implies $P\simeq 2h a
T^4/3$, where $a$ is black body constant and $T$ denotes the
midplane temperature of the disk. The energy equation is written
based on the second assumption. The last term of equation
(\ref{eq:rcom}) represents the height integrated radial magnetic
force which can be written as (LRN)
\begin{displaymath}
\int F_{\rm R}^{\rm mag} dz=\frac{1}{2\pi}(B_{\rm R} B_{\rm
z})_{\rm h}-\frac{1}{4\pi R^2}\frac{d}{d R}[h
R^{2}<B_{\varphi}^{2}-B_{\rm R}^2>]
\end{displaymath}
\begin{equation}
-\frac{1}{4\pi}\frac{d}{d R}[h <B_{\rm
z}^{2}>]+\frac{1}{4\pi}\frac{dh}{dR}(B_{\varphi}^{2}+B_{\rm z
}^{2}-B_{\rm R}^{2})_{\rm h},
\end{equation}
where $<\cdots>\equiv\int_{-h}^{h}dz(\cdots)/(2h)$, and the $h$
subscript denotes that the quantity is evaluated at the upper
disk plane, i.e. $z=h$. Similarly, integration over $z$ of the
azimuthal  equation of motion gives

\begin{equation}
R\Sigma v_{\rm R}\frac{d}{d R}(Rv_{\varphi})=\frac{d}{d
R}[R^{3}\nu\Sigma\frac{d}{d R}(\frac{v_{\varphi}}{R})]+\int
F_{\varphi}^{\rm mag} dz,\label{eq:phicom}
\end{equation}
where
\begin{displaymath}
\int F_{\varphi}^{\rm mag}
dz=\frac{1}{2\pi}(R^{2}B_{\varphi}B_{\rm z})_{\rm
h}-\frac{1}{2\pi}\frac{dh}{dR} (R^{2}B_{\rm R}B_{\varphi})_{\rm h}
\end{displaymath}
\begin{equation}
+\frac{1}{2\pi}\frac{d}{dR}[hR^{2} <B_{\rm R}B_{\varphi}>].
\end{equation}
Here, the last term of equation (\ref{eq:phicom}) represents the
height integrated toroidal component of magnetic force. Note that
this form of azimuthal equation of motion is not exactly similar
to the original slim disk model (e.g., Abramowicz et al. 1988),
in which  the well-know $\alpha p$  prescription of Shakura \&
Sunyaev (1973) has been used as a general approximate form for
the  $\varphi r$ component of the viscous stress tensor
($\tau_{\varphi\rm r}=-\alpha p$). Here, we replaced $\alpha p$
prescription by a diffusive viscosity prescription, i.e.
\begin{equation}
\tau_{\varphi\rm r}=\rho\nu r\frac{\partial\Omega}{\partial r},
\end{equation}
where $\rho$ is the density, $\nu$ is a kinematic viscosity
coefficient, and $\Omega$ is the angular velocity of matter in
the disk. The above prescription leads to equation
(\ref{eq:phicom}). In our model, we also assume
\begin{equation}
\nu=\alpha c_{\rm s} h,\label{eq:vis}
\end{equation}
where $c_{\rm s}$ is the local sound speed and $\alpha$ is a
constant less than unity.

The $z$ component of equation of motion gives the condition for
vertical hydrostatic balance, which can be written as (Lovelace
et al. 1987)
\begin{equation}
(\frac{h}{R})^{2}+q(\frac{h}{R})-\frac{2P}{\Sigma v_{\rm
K}^2}=0,\label{eq:zcom}
\end{equation}
where $v_{\rm K}=\sqrt{GM_{\ast}/R}$ is the Keplerian velocity and
\begin{equation}
q=\frac{R[(B_{\varphi})_{\rm h}^{2}+(B_{\rm R})_{\rm
h}^{2}]}{4\pi\Sigma v_{\rm K}^2}.
\end{equation}

Now, we can treat the internal magnetic field using the induction
equation. LRN showed that the variation of $B_{\rm z}$ with $z$
within the disk is negligible for even field symmetry. Moreover,
$B_{\rm R}$ and $B_{\varphi}$ are odd functions of $z$ and
consequently $\partial B_{\rm R}/\partial z\approx (B_{\rm
R})_{\rm h}/h$ and $\partial B_{\varphi}/\partial z\approx
(B_{\varphi})_{\rm h}/h$. Thus,
\begin{displaymath}
B_{\rm R}(R,z)=\frac{z}{h}(B_{\rm R})_{\rm h},
B_{\varphi}(R,z)=\frac{z}{h}(B_{\varphi})_{\rm h},
\end{displaymath}
\begin{equation}
 B_{\rm z}(R,z)=B_{\rm z}(R),
\end{equation}
and the induction equation reads
\begin{equation}
-RB_{\rm z} v_{\rm R}-\frac{\eta R}{h}(B_{\rm R})_{\rm h}+\eta
R\frac{d B_{\rm z}}{d R}=0,\label{eq:induction}
\end{equation}
where the magnetic diffusivity $\eta$ has the same units as
kinematic viscosity. We assume that the magnitude of $\eta$ is
comparable to that of the turbulent viscosity $\nu$ (e.g.,
Bisnovatyi-Kogan \& Ruzmaikin 1976; Shadmehri 2004). Exactly in
analogy to alpha prescription for $\nu$, we are using a similar
form for the magnetic diffusivity $\eta$,
\begin{equation}
\eta=\eta_{0}c_{\rm s}h,
\end{equation}
where $\eta_{0}$ is a constant. Note that $\eta$ is {\it not}
constant and depends on the physical variables of the flow, and in
our self-similar solutions, as we will show, $\eta$ scales with
radius as a power law. This form of scaling for diffusivity has
been widely used by many authors (e.g., Lovelace, Wang \& Sulkanen
1987; LRN ; Ogilvie \& Livio 2001; R\"{u}diger \& Shalybkov 2002).

While equation (\ref{eq:induction}) describes transport of a
large-scale magnetic field (here, $B_{\rm z}(R)$), the values of
$(B_{\rm R})_{\rm h}$ and $(B_{\varphi})_{\rm h}$ are determined
by the filed solutions external to the disk. Instead, we are
following  approach of LNR, in which the external field solutions
obey the relations
\begin{equation}
(B_{\rm R})_{\rm h}=\beta_{\rm r} B_{\rm z}, (B_{\varphi})_{\rm
h}=\beta_{\varphi} B_{\rm z},
\end{equation}
where $\beta_{\rm r}$ and $\beta_{\varphi}$ are constants of
order unity ($\beta_{\varphi}<0$). Thus, one can simply show that
$<B_{\rm R}^{2}>=\beta_{\rm r}^{2} B_{\rm z}^{2}/3$,
$<B_{\varphi}^{2}>=\beta_{\varphi}^{2} B_{\rm z}^{2}/3$ and
$<B_{\rm R}B_{\varphi}>=\beta_{\rm r}\beta_{\varphi}B_{\rm
z}^{2}/3$.

To close the equations of our model, we can write the energy
equation describing the thermal state of the flow as
\begin{equation}
\rho T v_{\rm R}\frac{dS}{dR}=Q_{\rm vis}+Q_{\rm
Joule},\label{eq:energy}
\end{equation}
where $S$ is the specific entropy (per unit mass) and $T$ is
midplane temperature of the disk. For the heating term, we may
have two sources of dissipation: the viscous and resistive
dissipations due to a turbulence cascade. So, $Q_{\rm vis}$ and
$Q_{\rm Joule}$ represent viscous dissipation due to the radial
motion and the Joule heating rate, respectively,
\begin{equation}
Q_{\rm vis}=\rho\nu R^{2}(\frac{d\Omega}{dR})^{2},
\end{equation}
and
\begin{equation}
Q_{\rm Joule}=\frac{\eta}{4\pi h^{2}}[2(B_{\rm R})_{\rm
h}^{2}+\frac{3}{5}(B_{\varphi})_{\rm h}^{2}]
\end{equation}

Now we have constructed our model and the main equations of the
model are equations (\ref{eq:masscon}), (\ref{eq:rcom}),
(\ref{eq:phicom}), (\ref{eq:zcom}), (\ref{eq:induction}) and
(\ref{eq:energy}). In the next section, we will present
self-similar solutions of these equations.
\section{self-similar solutions}
The equations of our model are reduced to standard equations of
the slim disk, if we set all the magnetic terms equal to zero. As
we discussed, we are considering a radiation-dominated disk, in
which the gas pressure has been neglected comparing to the
radiation pressure. After some algebraic manipulations we get to
the following set of self-similar solutions:
\begin{equation}
\Sigma(R)=a \Sigma_{0}(\frac{R}{R_0})^{-1/2},
\end{equation}
\begin{equation}
v_{\varphi}(R)=b
\sqrt{\frac{GM_{\ast}}{R_0}}(\frac{R}{R_0})^{-1/2},
\end{equation}
\begin{equation}
v_{\rm R}(R)=- c
\sqrt{\frac{GM_{\ast}}{R_0}}(\frac{R}{R_0})^{-1/2},
\end{equation}
\begin{equation}
P(R)= d \frac{\Sigma_{0}GM_{\ast}}{R_0}(\frac{R}{R_0})^{-3/2},
\end{equation}
\begin{equation}
B_{\rm z}(R)=e
\sqrt{4\pi\Sigma_{0}\frac{GM_{\ast}}{R_{0}^{2}}}(\frac{R}{R_0})^{-5/4},
\end{equation}
\begin{equation}
h(R)=f R_{0} (\frac{R}{R_0}),
\end{equation}
where $\Sigma_{\rm 0}$ and $R_{\rm 0}$ provide convenient units
with which the equations can be written in non-dimensional form.
Thus, we obtain the following system of dimensionless equations,
to be solved for $a$, $b$, $c$, $d$, $e$ and $f$:

\begin{equation}
ac=\dot{m},
\end{equation}
\begin{equation}
-\frac{1}{2}ac^{2}-ab^{2}=\frac{3}{2}d-a+[2\beta_{\rm
r}+(\frac{5+\beta_{\varphi}^{2}-\beta_{\rm r}^{2}}{2})f]e^{2},
\end{equation}
\begin{equation}
-\frac{1}{2}abc=-\frac{3\alpha}{4}\sqrt{\frac{d}{a}}fab+\beta_{\varphi}(2-\beta_{\rm
r }f)e^{2},
\end{equation}
\begin{equation}
af^{2}+(\beta_{\varphi}^{2}+\beta_{\rm r}^{2})fe^{2}-2d=0,
\end{equation}
\begin{equation}
c-\eta_{0}\beta_{\rm
r}\sqrt{\frac{d}{a}}-\frac{5}{4}\eta_{0}f\sqrt{\frac{d}{a}}=0,
\end{equation}
\begin{equation}
\frac{1}{4}c\sqrt{ad}=\frac{3\alpha}{8}fab^{2}+\frac{1}{3}\eta_{0}(2\beta_{\rm
r }^{2}+\frac{3}{5}\beta_{\varphi}^{2})e^{2},
\end{equation}
where $\dot{m}=\dot{M}/(2\pi \Sigma_{0}\sqrt{GM_{\ast}R_{0}})$ is
nondimensional mass accretion rate. Thus, our input parameters
are  $\alpha$, $\eta_{0}$, $\beta_{\rm r}$, $\beta_{\varphi}$ and
$\dot{m}$. We will present values of the physical quantities in
non-dimensional form.

In the limit of nonmagnetic case, the above equations can be
solved analytically:
\begin{equation}
a=\frac{3\sqrt{2}\dot{m}\alpha}{\sqrt{25+36\alpha^{2}}-5}\simeq
\frac{5\sqrt{2}\dot{m}}{6\alpha},
\end{equation}
\begin{equation}
b=\frac{1}{3\alpha}\sqrt{\sqrt{25+36\alpha^{2}}-5}\simeq\sqrt{\frac{2}{5}},
\end{equation}
\begin{equation}
c=\frac{\sqrt{25+36\alpha^{2}}-5}{3\sqrt{2}\alpha}\simeq
\frac{6\alpha}{5\sqrt{2}},
\end{equation}
\begin{equation}
d=\frac{\sqrt{2}}{3\alpha}\dot{m},
\end{equation}
\begin{equation}
f=\frac{\sqrt{2}}{3\alpha}\sqrt{\sqrt{25+36\alpha^{2}}-5}\simeq\frac{2}{\sqrt{5}}.
\end{equation}
The second relation in each equation refers to the limit
$\alpha\ll 1$. The scaling of our solutions are different from
Wang \& Zhou (1999) who analyzed self-similar solution of
optically thick advection-dominated flows. Because they applied
$\alpha p$ prescription for viscous stress tensor, but a diffusive
prescription has been used in our model, i.e. equation
(\ref{eq:vis}).

The above nonmagnetic solutions show that the surface density
increases with accretion rate, and decreases inversely with
$\alpha$. However, the radial velocity is directly proportional
to the viscosity coefficient $\alpha$. The gas rotates  with
sub-Keplerian angular velocity, i.e.
$\Omega\approx\sqrt{\frac{2}{5}}\Omega_{\rm K}$. Note that except
for the surface density and the pressure, the other physical
quantities are independent of the accretion rate but depend only
on the viscosity coefficient $\alpha$. An interesting feature is
that  the opening angle of the disk is fixed $h/R\approx
2/\sqrt{5}$, independent of $\alpha$ and of the mass accretion
rate $\dot{m}$.

\begin{figure}
\epsfig{figure=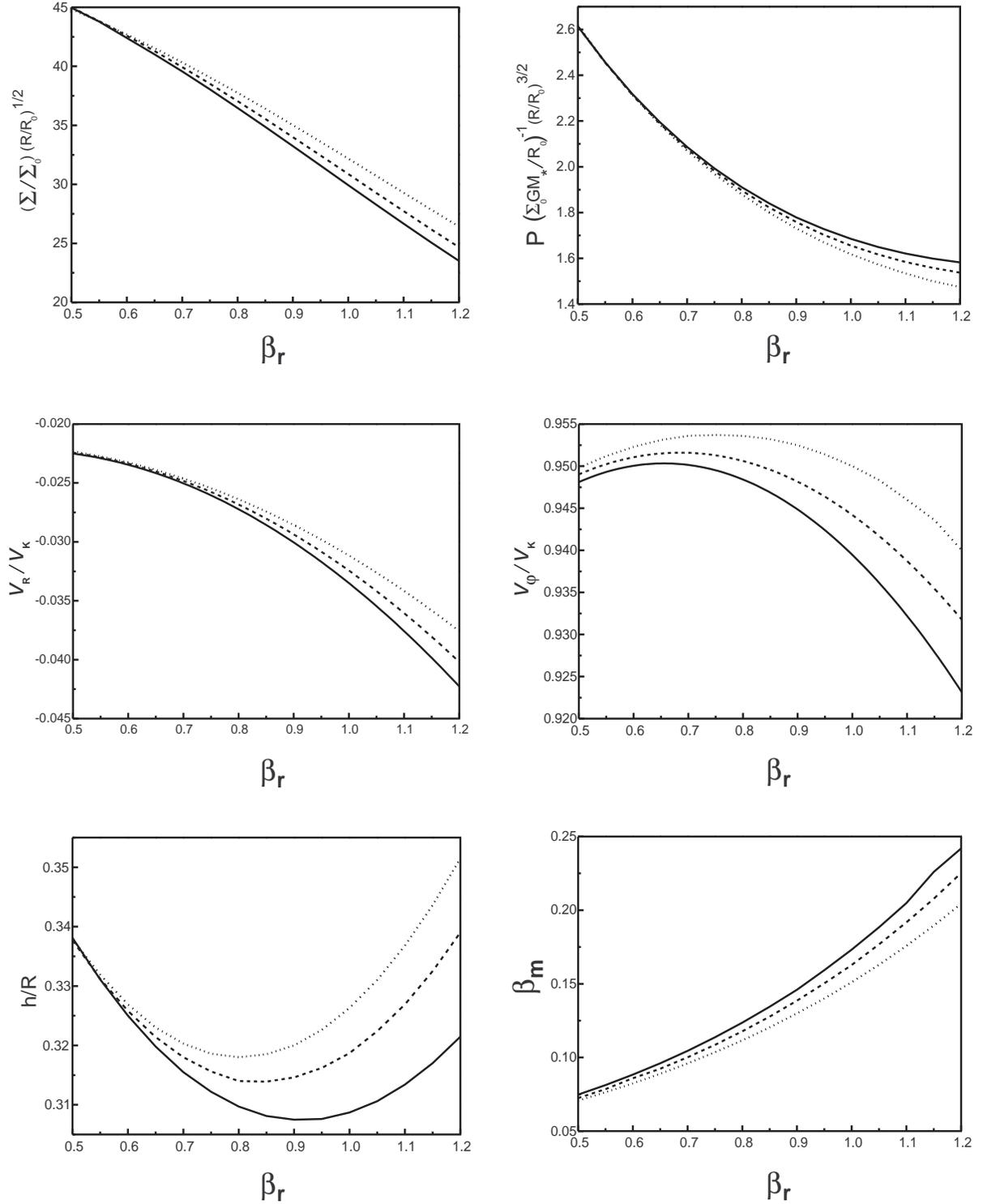,angle=0,width=\hsize} \caption{Profiles of
some height-averaged physical quantities for the disk, as a
function of $\beta_{\rm r}$, for $\dot{m}=1$, $\eta_0=0.1$,
$\alpha=0.01$ and $\beta_{\varphi}=-0.8$ (solid line),
$\beta_{\varphi}=-0.9$ (dashed line) and $\beta_{\varphi}=-1.1$
(dotted line). The surface density $\Sigma$, the radial velocity
$v_{\rm R}$, the rotational velocity $v_{\varphi}$, the pressure
$P$, half-thickness $h$ are drawn in dimensionless form. The
ratio of the magnetic pressure at the surface of the disk to the
pressure is presented by $\beta_{m}$.}\label{fig:figure1}
\end{figure}
\begin{figure}
\epsfig{figure=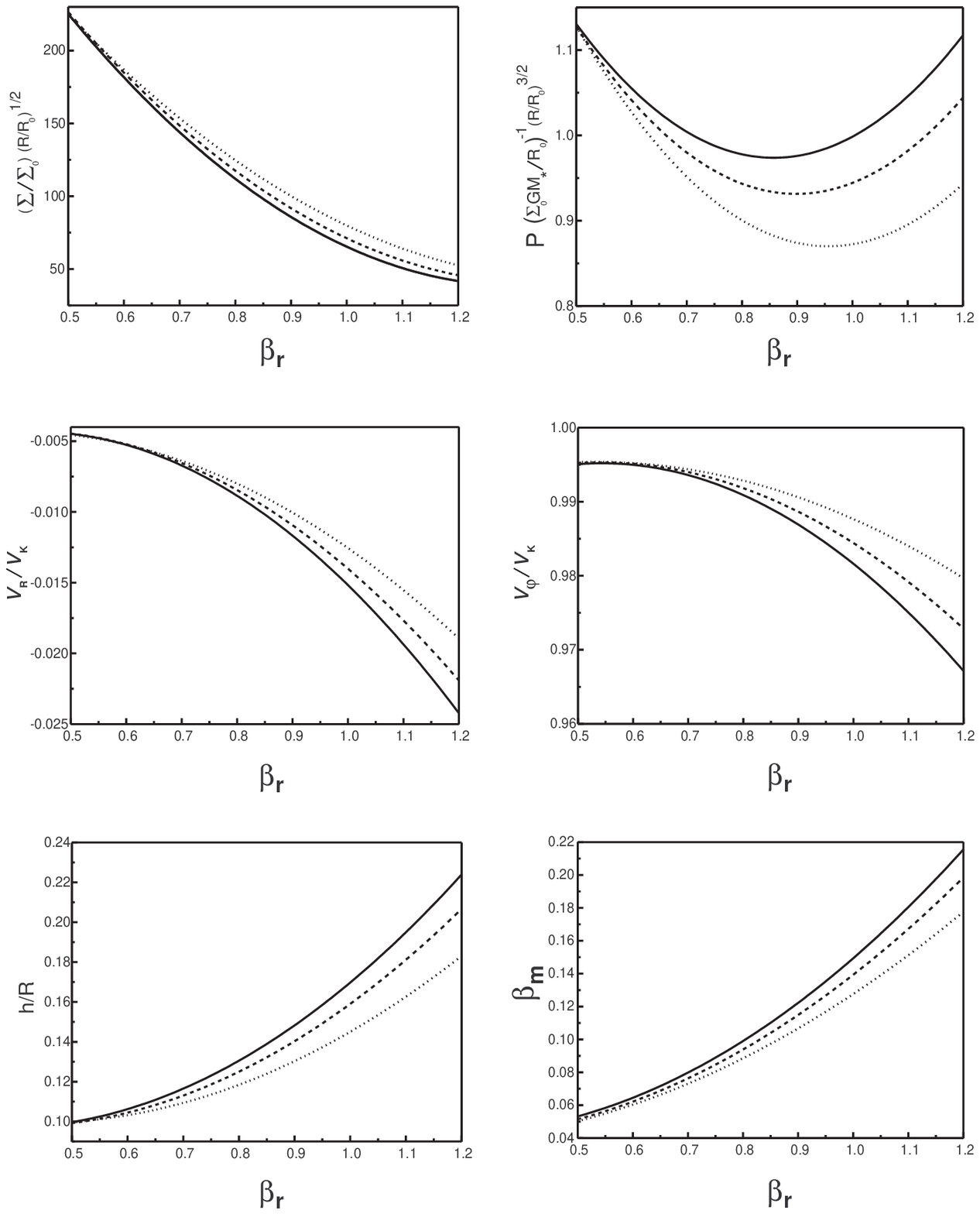,angle=0,width=\hsize} \caption{Same as
Figure 1, but in the case of $\alpha=0.001$, $\dot{m}=1$,
$\eta_0=0.1$. Solid, dashed, and dotted lines refer to
$\beta_{\varphi}=-0.8$, $\beta_{\varphi}=-0.9$  and
$\beta_{\varphi}=-1.1$.}\label{fig:figure2}
\end{figure}

Although the radial scaling of the solutions are similar to the
nonmagnetic case, we can see significant differences because of
the magnetic field effect. The first important effect of the
magnetic field on the disk structure is a squeezing effect, where
the scale height $h$ is reduced comparing to the nonmagnetic
case. In fact, the squeezing effect of the large-scale magnetic
field counterbalances the thickening of the disk generated by
advection. Figure 1 shows the behaviour of physical disk
quantities as a function of $\beta_{\rm r}$, for three different
values of $\beta_{\varphi}=-0.8, -0.9, -1.1$. The other input
parameters are assumed as $\dot{m}=1$, $\eta_0=0.01$,
$\alpha=0.01$. Generally, $\beta_{\rm r}$ and $\beta_{\varphi}$
are parameters of order unity, however, we consider
$\beta_{\varphi}$ around unity but changing $\beta_{\rm r}$ from
0.5 to 1.2.  For these input parameters, the surface density
significantly reduces from $a\simeq 117$ to a value between 25
and 45 depending on the magnetic field configuration. However, as
$\beta_{\varphi}$ increases, the surface density slightly
increases for a fixed $\beta_{\rm r}$. The rotation velocity is
below Keplerian and as $\beta_{\rm r}$ increases, the rotation
rate reaches to a maximum and then decreases. Also, the radial
velocity increases with $\beta_{\rm r}$ or $\beta_{\varphi}$,
however, is significantly below free-fall velocity. But comparing
to the nonmagnetic solution, the magnetic field tends to increase
the radial velocity. For example, the above input parameters gives
$c=8.48\times 10^{-3}$ for nonmagnetic flow, but the field causes
$c$ increases to a value between $0.02$ to $0.04$. Magnetic field
causes the opening angle of the disk decreases.

Figure 2 is the same as Figure 1, but with lower viscosity
coefficient, i.e. $\alpha=0.001$. We see this decrease of
$\alpha$ causes the surface density increases. While the radial
velocity decreases with $\alpha$, the rotation velocity becomes
closer to the Keplerian rate. We see that the opening angle of
disk for $\alpha=0.001$ is smaller than $\alpha=0.01$. Generally,
the physical quantities are sensitive to the viscosity coefficient
$\alpha$.

We repeated the above calculations for $\alpha=\eta_0=0.01$ and
found that the solutions weakly change because of the  variations
of $\beta_{\rm r}$ and $\beta_{\varphi}$. In this case, the
solutions are qualitatively similar to Figures 1 and 2, but the
scaling is somewhat different. For example, for these input
parameters, we find the ratio of $v_{\varphi}/v_{\rm K}$ between
$0.66$ and $0.69$ depending on $\beta_{\rm r}$ and
$\beta_{\varphi}$. Regarding to Figure 1 which is for
$\eta_0=0.1$, we can say as the resisitivity coefficient $\eta_0$
decreases, the dependence of solutions on the outside field
solutions becomes weaker. The nonmagnetic solutions show that the
rotational and the radial velocity and the opening angle are
independent of the mass accretion rate. This result is valid even
in magnetic case, as we found by changing the accretion rate and
keeping other input parameters constant.

\section{conclusion}
Although investigating the behavior of the magnetic field and
associated currents within the disk was not the main purpose of
our study, we studied the effect of a large-scale magnetic field
with open field lines on the structure of an optically thick
accretion disk. While the field outside of the disk  treated in a
phenomenological way, we solved the height-averaged MHD equations
self-consistently using similarity technique in analogy to the
original study of optically thin ADAF by Narayan \& Yi (1994).
Our self-similar solutions reduce to the nonmagnetic solutions of
Wang \& Zhou (1999) for optically  thick advection-dominated
accretion flow. The disk structure and the field geometry are
closely linked. The magnetic field is dragged by the accreting
flow, however, the field tends to squeeze the disk and to
increase the radial velocity. The angular velocity of the flow is
less than the local Keplerian angular velocity by a factor which
depends on the magnetic field configuration.

Our simple self-similar solutions show that the effect of the
magnetic field can not be ignored in a realistic accretion model.
The present solutions may be applied to the X-ray galactic and
extragalactic sources, when the accretion rate is high and
radiation dominated regime takes place. The emergent spectrum of
such a disk can be calculated using our magnetized self-similar
solutions and considering the energy transfer in the vertical
direction. We will discuss this problem in future.

\end{document}